\documentclass{ifacconf}
\usepackage{caption}
\usepackage{subcaption}
\usepackage{float}
\usepackage{amsmath}
\usepackage{amssymb}
\usepackage{algorithmic}
\usepackage{float}
\usepackage{graphicx}      
\usepackage{natbib}        
\usepackage{MnSymbol} 
\usepackage{bm}
\usepackage{siunitx} 
\newcommand{\vect}[1]{\bm{#1}}

\usepackage[most]{tcolorbox}
\usepackage{tabularx}
\usepackage{booktabs}
\usepackage{siunitx}
\usepackage{listings}
\usepackage{microtype}

\tcbset{
  width=\columnwidth,
  breakable,                
  enhanced,
  boxrule=0.3pt,
  colframe=black!40,
  colback=gray!3,
  left=1mm,right=1mm,top=0.8mm,bottom=0.8mm,
  before skip=5pt, after skip=5pt
}
\DeclareRobustCommand\dashed{\tikz[baseline=-0.6ex]\draw[thick,dashed] (0,0)--(0.54,0);} 

\usepackage{tcolorbox}
\newtcolorbox{llmprompt}[1][]{
  colback=gray!10,
  colframe=black!70,
  boxrule=0.5pt,
  arc=1pt,
  left=6pt,
  right=6pt,
  top=6pt,
  bottom=6pt,
  fonttitle=\bfseries,
  title=#1
}

\newcommand{\matr}[1]{\mathbf{#1}} 
\newtcolorbox{promptbox}{title=\textbf{Prompt}, colback=blue!3}
\newtcolorbox{toolbox}{title=\textbf{Tool use (queries \& calls)}, colback=green!5}
\newtcolorbox{evidencebox}{title=\textbf{Retrieved evidence (sample)}, colback=orange!5}
\newtcolorbox{decisionbox}{title=\textbf{Controller decision}, colback=purple!5}
\newtcolorbox{rationalebox}{title=\textbf{Brief rationale}, colback=purple!5}

\lstdefinestyle{sqlstyle}{
  basicstyle=\ttfamily\footnotesize,
  breaklines=true,
  breakatwhitespace=false,
  columns=fullflexible,
  keepspaces=true,
  frame=none
}

\begin{document}
\begin{frontmatter}

\title{Large Language Models for Control} 


\author[First,Second]{Adil Rasheed} 
\author[First]{Oscar Ravik}
\author[Third]{Omer San}

\address[First]{Department of Engineering Cybernetics, Norwegian University of Science and Technology, O. S. Bragstads plass 2, Trondheim, NO-7034, Norway (e-mail: adil.rasheed@ntnu.no)}
\address[Second]{Mathematics and Cybernetics, SINTEF Digital, Strindvegen 4, 7034 Trondheim, Norway.}
\address[Third]{Department of Mechanical and Aerospace Engineering, University of Tennessee, Knoxville, 37966 TN, USA.}

\begin{abstract} This paper investigates using large language models (LLMs) to generate control actions directly, without requiring control-engineering expertise or hand-tuned algorithms. We implement several variants: (i) prompt-only, (ii) tool-assisted with access to historical data, and (iii) prediction-assisted using learned or simple models to score candidate actions. We compare them on tracking accuracy and actuation effort, with and without a prompt that requests lower actuator usage. Results show prompt-only LLMs already produce viable control, while tool-augmented versions adapt better to changing objectives but can be more sensitive to constraints, supporting LLM-in-the-loop control for evolving cyber-physical systems today and operator and human inputs.
\end{abstract}

\begin{keyword}
Large Language Model, Hybird Analysis and Modeling, Control.
\end{keyword}

\end{frontmatter}
\section{Introduction}
\label{sec:introduction}
Designing feedback controllers for real systems remains resource-intensive and requires specialized expertise. Even for relatively simple processes, engineers must select or identify a model, encode objectives and constraints, tune gains or penalties, and then validate performance under disturbances and uncertainty \citep{AstromMurray2010,RawlingsMayneDiehl2009,Mayne2014}. Any change in operating conditions, equipment aging, sensor upgrades, or the introduction of new objectives typically triggers another round of modeling and retuning. In practice, once a controller is deployed, it rarely continues to exploit the growing body of domain knowledge found in logs, documentation, standards, and the wider technical literature; it becomes a fixed artifact in a world that keeps changing.

Large language models (LLMs) such as ChatGPT \citep{OpenAI2024g4t1}, LLaMA \citep{Touvron2023loa}, and Gemini \citep{Team2025gaf1} offer a complementary path. They can express control intent in plain language, interpret heterogeneous contexts (telemetry, tables, documentation), and when paired with tools such as databases, simulators, or learned predictors, can reason about candidate actions without demanding deep control expertise from the operator \citep{Yao2023ReAct,Schick2023Toolformer,Cai2024LLMToolMakers}. Because LLMs can be refreshed with new information, they also provide a mechanism for controllers to evolve with the surrounding knowledge ecosystem: updated procedures, best practices, or even public web resources can be incorporated without rewriting the control algorithm from scratch. This aligns well with data-centric infrastructures such as digital twins \citep{Rasheed2020dtv}, where up-to-date process information is continuously available for decision-making.

This paper explores the potential of LLMs as controllers. We study architectures in which an LLM receives high level goals, consults historical data or prediction models as tools, and outputs actuator commands together with short rationales. Our central hypothesis is that such \emph{reasoning-in-the-loop} systems can reduce the expertise and effort required to build and maintain controllers while remaining adaptable as new knowledge becomes available. To show that this is not only a conceptual proposal, we implement the different LLM variants on a physical greenhouse testbed with sensing, actuation, and a digital-twin backend, and we report closed-loop results that compare prompt-only, SQL-assisted, and prediction-assisted controllers under the same operating conditions.

The remainder of the paper is organised as follows. Section~\ref{sec:methodology} describes the methodology for building and prompting LLM-based controllers with optional tool assistance. Section~\ref{sec:resultsanddiscussions} presents and discusses the experimental results. Section~\ref{sec:conclusion} concludes the paper and outlines directions for future research.

\section{Methodology}
\label{sec:methodology}
In this section we briefly describe the experimental setup, the predictive models, the LLM-based controllers, and the dataset used for training.

\subsection{Experimental setup}
\label{subsec:expsetup}
\begin{figure}
    \centering
    \includegraphics[width=\linewidth]{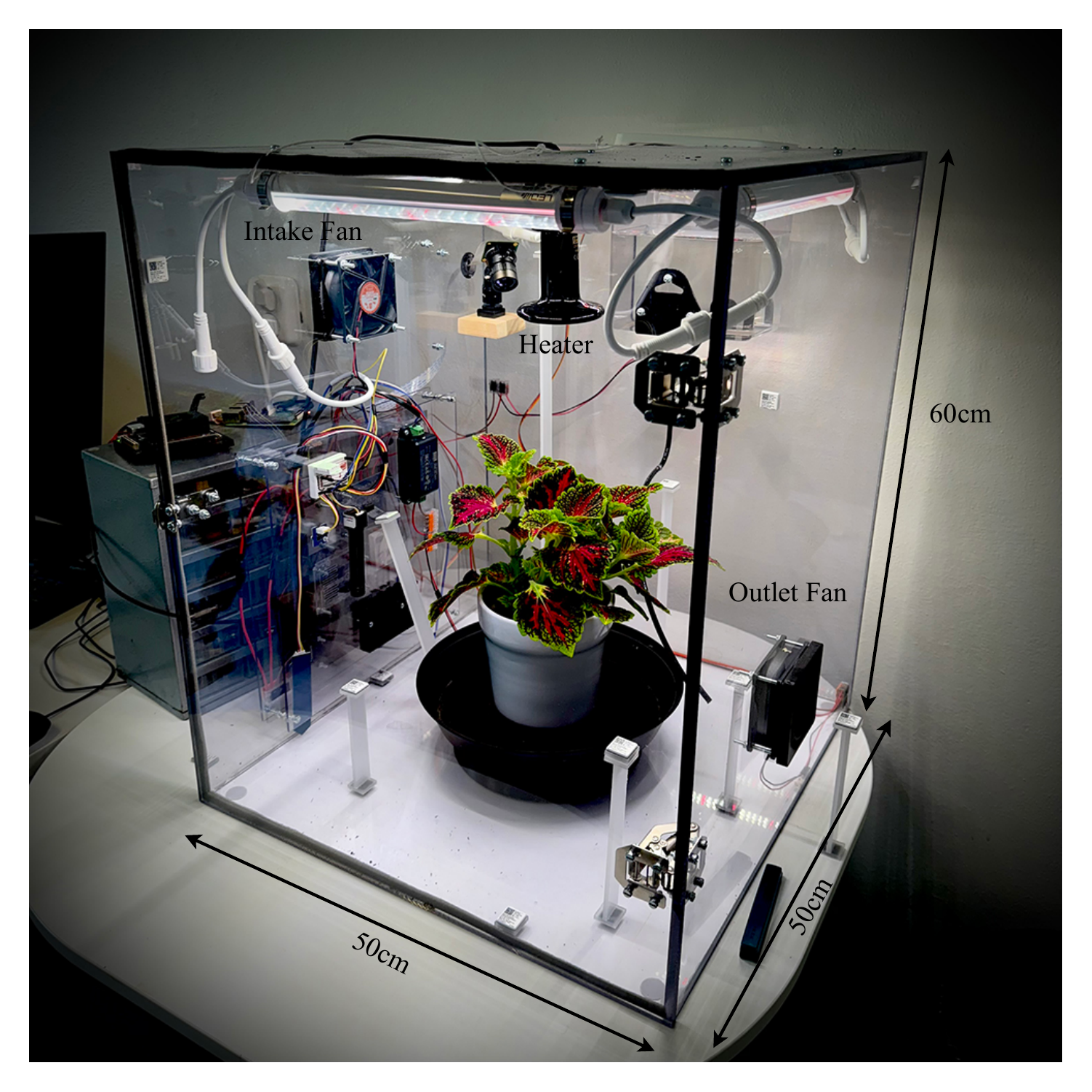}
    \caption{Asset}
    \label{fig:GreenHouseAsset}
\end{figure}
The experimental platform employed in this study (Fig.~\ref{fig:GreenHouseAsset}) consists of a compact greenhouse designed to emulate a controlled environment. The enclosure forms a cuboidal structure with internal dimensions of $50\text{cm} \times 50\text{cm} \times 60\text{cm}$, fabricated from $8\text{mm}$ acrylic sheet. Environmental regulation within the greenhouse is accomplished via a $H_{max}=100\text{W}$ ceramic heater for thermal management, full-spectrum LED luminaires optimized for plant growth, and high-capacity intake and exhaust fans (maximum airflow rate $F_{max} = 68\text{m}^3/\text{h}$) for ventilation and CO$_2$ exchange. Irrigation is automated through a solenoid valve connected to an external water reservoir.

A comprehensive sensor array continuously monitors key environmental variables, including air temperature, relative humidity, CO$_2$ concentration, light intensity, soil moisture, and water-tank level. Although the primary focus of this work is temperature regulation through coordinated fan and heater control, the inclusion of a live plant introduces biological variability into the microclimate. The plant’s gradual physiological responses act as a natural source of uncertainty, thereby providing a realistic context for evaluating the robustness and adaptability of the proposed control framework under dynamic, real-world conditions.

\subsection{Predictive modeling}
\label{subsec:predictivemodeling}
\subsubsection{Hybrid Analysis and Modeling:} The HAM approach used in this work is called the COrrective Source Term Approach (CoSTA), whose theoretical foundation can be found in \cite{Blakseth2022cpb,Blakseth2022dnn}. It augments a physics based model (PBM) (given by Equation~\ref{eq:method-hybrid-model} without the source term $r$) with a data-driven correction term $r(T, T_{amb}, H, F; \boldsymbol{\theta})$ (see Equation~\ref{eq:method-hybrid-model}) to compensate for the modeling assumptions described in Section~\ref{subsec:expsetup}. 
\begin{equation}
\frac{d T}{dt} = \frac{H}{\rho V C_p} - \frac{F (T - T_{amb})}{V} + r(T, T_{amb}, H, F; \boldsymbol{\theta}) 
\label{eq:method-hybrid-model}
\end{equation}
where the term on the left hand side represents the change in inside temperature with respect to time, while on the right hand side, the first term represents the heat input from the heater, where $H$ is the heating power, $\rho$ is the air density, $V$ is the enclosure volume, and $C_p$ is the specific heat capacity of air. Since the heater is operated using a duty cycle $u_h$, $H$ is expressed as $u_{h}H_{max} $. The $u_h$ can be changed in discrete steps of 0.05 from 0 to 1. The second term represents the heat exchange induced by the fan, where the airflow is modeled as $F = u_{f}F_{max}$, with $u_f$ denoting the fan’s ON/OFF state. $T_{amb}$ is the external ambient temperature. 
\begin{table}
\centering
    \caption{Model parameter for computing the corrective term in the HAM model}
    \label{tab:method-model-parameters-hybrid}
    \begin{tabular}{l l | l l}
        \toprule
        \textbf{Parameter} & \textbf{Value} & \textbf{Layer} & \textbf{Description}\\
        \midrule
        Epochs & 1000 & Linear 1       & Linear\\
        Optimizer & Adam & ReLU 1   & ReLU\\
        Learning rate & 0.001 & Dropout 1      & Dropout(p=0.2)\\
        Loss function & MSE & Linear 2       & Linear\\
        Batch size & 64 & ReLU 2   & ReLU\\
        Hidden size & 64 & Dropout 2      & Dropout(p=0.2)\\
        Min delta & $5 e^{-4}$ & Linear 3       & Linear\\
        Tolerance & 10 & &\\
        \bottomrule
    \end{tabular}
\end{table}
The source term $r(T, T_{amb}, Q, F; \boldsymbol{\theta})$ is a data-driven learned component that accounts for the modeling error due to the uncertainty in the input parameters and the simplifying assumptions made regarding the dynamics of the setup. Here, $\boldsymbol{\theta}$ denotes the learnable parameters of the DDM used to represent the residual. For the HAM model (CoSTA), Equation~\ref{eq:method-hybrid-model} is first solved  without the correction term to obtain an uncorrected temperature $\hat{T}$ at the next time step. The variables $\hat{T}$, $T_{amb}$, $H$, and $F$ are then used by a neural network to compute the corrective source term $r$. Subsequently, Equation~\ref{eq:method-hybrid-model} is solved again with this corrective term to obtain the correct temperature $T_{t+1}$. Details regarding the hyperparameters of the neural network used to learn this source term is given in Table~\ref{tab:method-model-parameters-hybrid}. 

\subsubsection{Linear Model}
\label{subsubsec:theory-lm}

The system is modelled as a linear autoregressive model with exogenous inputs (ARX), in which the future temperature depends on a finite history of past temperatures and past control inputs. The model is written as
\begin{align}
T_{t+1} &= a_1 T_t + a_2 T_{t-1} + \dots + a_p T_{t-p+1} \notag \\
        &\quad + \sum_{j=1}^{q} b_{j}^{(h)} u_{h,t-j+1} 
              + \sum_{j=1}^{q} b_{j}^{(f)} u_{f,t-j+1},
\label{eq:method-arx-equation}
\end{align}
where $a_i$ are the autoregressive coefficients, and $b_{j}^{(h)}$ and $b_{j}^{(f)}$ capture the influence of the heater duty cycle $u_h$ and the fan state $u_f$, respectively. The look-back horizon $p$ captures the temporal dependence of temperature on its own past, while $q$ captures the effect of past control actions. This linear ARX structure enables the controller to predict the temperature evolution from recent measurements, together with the applied control inputs, and it provides a lightweight alternative to the LSTM and HAM predictors.

\subsubsection{Long Short Term Memory}
To capture nonlinear dynamics and long-range dependencies in the system, a data-driven approach based on LSTM \citep{Hochreiter1997lsta} networks is employed. At each time step $t$, the LSTM updates its hidden state $\vect{h_t}$ and cell state $\vect{c_t}$ according to the following equations:  
\begin{align}
    \vect{f_t} &= \sigma \left( \matr{W_f} [\vect{h_{t-1}}, \vect{x_t}] + \vect{b_f} \right) \\
    \vect{i_t} &= \sigma \left( \matr{W_i} [\vect{h_{t-1}}, \vect{x_t}] + \vect{b_i} \right) \\
    \tilde{\vect{c_t}} &= \tanh \left( \matr{W_c} [\vect{h_{t-1}}, \vect{x_t}] + \vect{b_c} \right) \\
    \vect{c_t} &= \vect{f_t} \odot \vect{c_{t-1}} + \vect{i_t} \odot \tilde{\vect{c_t}} \\
    \vect{o_t} &= \sigma \left( \matr{W_o} [\vect{h_{t-1}}, \vect{x_t}] + \vect{b_o} \right) \\
    \vect{h_t} &= \vect{o_t} \odot \tanh(\vect{c_t})
\end{align}  
where $\sigma(\cdot)$ is the sigmoid activation function, $\odot$ denotes element-wise multiplication, $\vect{f_t}$ is the forget gate, $\vect{i_t}$ the input gate, $\vect{o_t}$ the output gate, and $\tilde{\vect{c_t}}$ the candidate cell state.   In this application, the input sequence $\vect{x_t}$ consists of past temperature values and control signals ($u_h$ and $u_f$), while the output is the predicted temperature at the next time step. By exploiting its memory structure, the LSTM can model nonlinear interactions and capture temporal dependencies over longer horizons compared to the linear ARX model. The hyperparameters of the LSTM model used is provided in the Table~\ref{tab:method-model-parameters-lstm}.
\begin{table}
\centering
    \caption{Model parameters for the LSTM model.}
    \label{tab:method-model-parameters-lstm}
    \begin{tabular}{l l | l l}
        \toprule
        \textbf{Parameter} & \textbf{Value} & \textbf{Layer} & \textbf{Description}\\
        \midrule
        Epochs & 5000 & LSTM 1         & LSTM\\
        Optimizer & Adam & Linear 1           & Linear\\
        Learning rate & 0.001 & Dropout 1      & Dropout(p=0.2)\\
        Loss function & MSE & LSTM 2         & LSTM\\
        Batch size & 40 & Linear 2           & Linear\\
        Hidden size & 64 & Dropout 2      & Dropout(p=0.2)\\
        Min delta & $5 e^{-4}$ & LSTM 3         & LSTM\\
        Tolerance & 10 & Linear 3           & Linear\\
        \bottomrule
    \end{tabular}
\end{table}

\subsection{Data generation}
\label{subsec:datageneration}
The data used for training and evaluating the models were generated using the experimental setup described in Section~\ref{subsec:expsetup}, in which predefined input sequences of heater duty cycle ($u_h$) and fan ($u_f$) input were applied to the system. The resulting temperature ($T$) was recorded with a sampling interval of $60\text{s}$. 15 time series, each lasting an average of 212 minutes, were created. 
For the  training of the predictive models, the time series were transformed into features and labels. This transformation was performed in the same way for the linear and LSTM models, whereas the HAM model required a different preparation step. To prepare the data for LSTM / Linear model training, the time series were segmented into overlapping feature-label pairs using a sliding window. Each feature vector contained the 10 most recent states and control inputs, and the corresponding label was the next state of the system. For the HAM model, the same raw signals $(T, T_{amb}, u_h, u_f)$ were reused, but the dataset was constructed to learn the corrective source term. First, the physics-based part of the hybrid model was advanced one step without correction using the measured inputs and ambient temperature, which produced an intermediate prediction $\hat{T}_{t+1}$. This uncorrected prediction, together with $T_{amb,t}$, $u_{h,t}$, and $u_{f,t}$, was fed as the input to the neural network, while the computed residual $r_{t+1}$ served as the target. 

\subsection{LLM-based controller}
\begin{figure}[htbp]
    \centering
    \begin{subfigure}{0.31\linewidth}
        \includegraphics[width=1\linewidth]{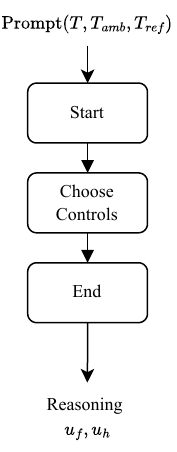}
        \caption{Simple model}
        \label{subfig:LLMArchitecture}
    \end{subfigure}
    \begin{subfigure}{0.33\linewidth}
        \includegraphics[width=1\linewidth]{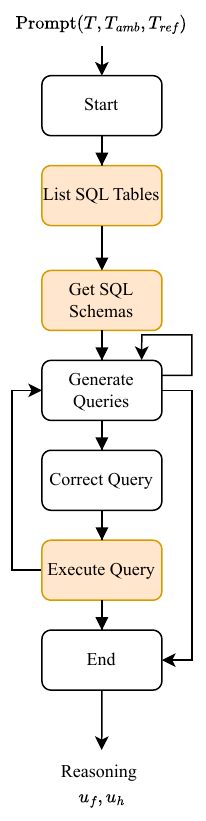}
        \caption{SQL query}
        \label{subfig:LLMArchitectureSQL}
    \end{subfigure}
    \begin{subfigure}{0.31\linewidth}
        \includegraphics[width=1\linewidth]{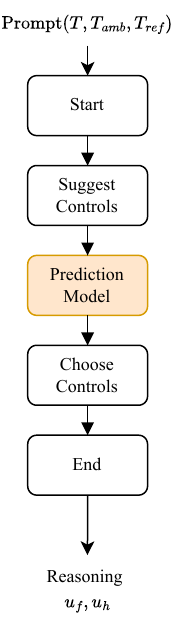}
        \caption{Prediction model}
        \label{subfig:LLMArchitecturePrediction}
    \end{subfigure}
    \caption{The architechture of the three different LLM controller implementations.}
    \label{fig:method-llm-graph}
\end{figure}

The four variants of LLM-based controllers for regulating temperature evolution were implemented using the Python library \textit{LangChain} \citep{Chase2022l} and OpenAI’s GPT-4o \citep{hg4}. \textit{LangChain} enables the construction of tool-augmented agents (e.g., predictive models, SQL databases) capable of executing complex tasks. Each controller outputs $u_f$ and $u_h$, along with an accompanying rationale that explains the selected values. This rationale provides transparency into the controller’s decision-making process.The different implementations are shown in Fig.~\ref{fig:method-llm-graph}. The examples of prompts that are sent as input to the controller are as follows:

\begin{llmprompt}[LLM prompt without control penalty]
    What should the control values heater\_duty\_cycle and fan\_on be set to in order to maintain a temperature of \textit{target temperature} degrees? The temperature now is \textit{current temperature} and the ambient temperature is \textit{ambient temperature} degrees. 
    It is important that the temperature in the greenhouse matches the target temperature exactly. Use \textit{tool}.
\end{llmprompt}

\begin{llmprompt}[LLM prompt with control penalty]
    What should the control values heater\_duty\_cycle and fan\_on be set to in order to maintain a temperature of \textit{target temperature} degrees? The temperature now is \textit{current temperature} and the ambient temperature is \textit{ambient temperature} degrees.
    The second priority is to match the target temperature accurately and the first priority is to have a minimal usage of the fan. Use \textit{tool}.
\end{llmprompt}

When the prompt is presented to the model, the placeholders \textit{target temperature}, \textit{current temperature}, \textit{ambient temperature}, and \textit{tool} are replaced with the corresponding numerical values of the target, current, ambient temperatures, and the prediction model / SQL database.

The first implementation is a controller that uses the LLM without any tools and obtains the controls through a single interaction with the LLM agent. The architecture of the model is shown in Fig.~\ref{subfig:LLMArchitecture}. The second implementation (Fig.~\ref{subfig:LLMArchitectureSQL}) uses archived historical data to suggest the best possible controls. The third implementation (Fig.~\ref{subfig:LLMArchitecturePrediction}) uses predictive models (Linear, HAM and LSTM) to simulate the next timesteps when a set of suggested controls is applied to the prediction model. The LLM/HAM models internally use multiple sets of controls, simulate the next timesteps using the prediction model, and, in the end, use the simulation results to determine the best possible control inputs. A penalty for control usage can be introduced by altering the agent's prompt and instructing it to reduce the actuation.  

\section{Results and discussions}
\label{sec:resultsanddiscussions}
\begin{figure}
    \centering
    \includegraphics[width=\linewidth]{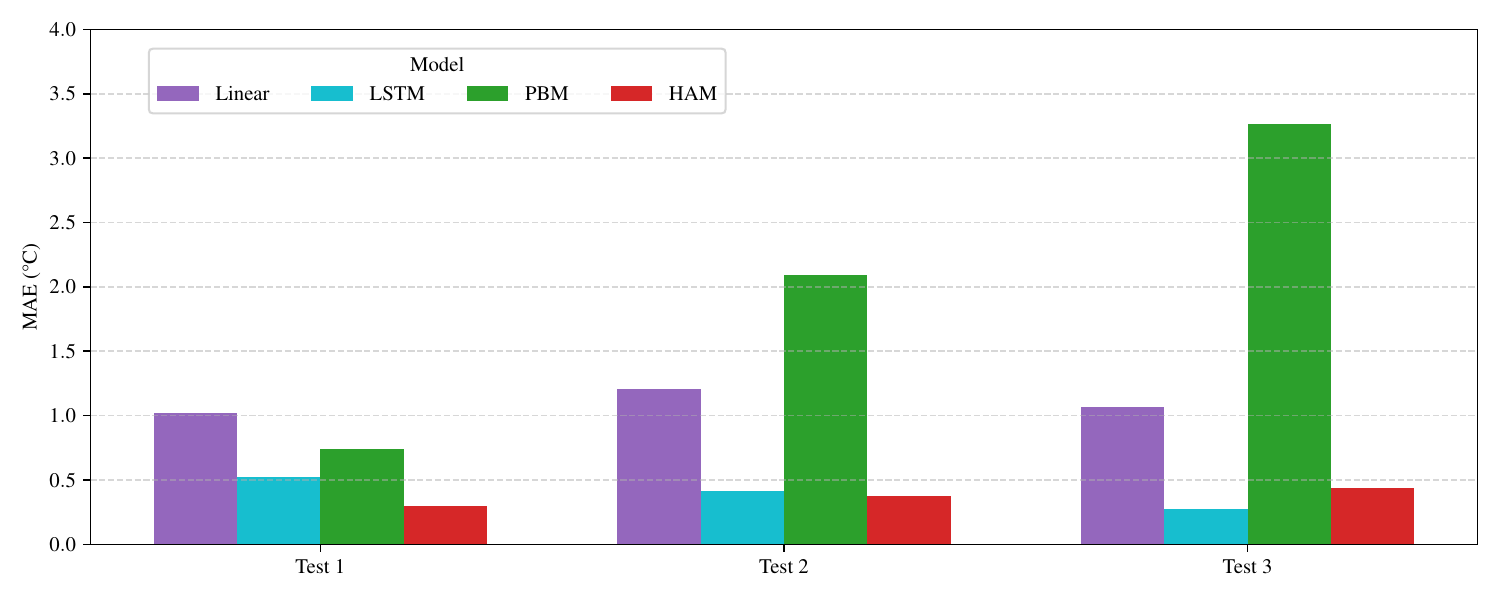}
    \caption{Inter-comparison of different modeles}
    \label{fig:modellingerror}
\end{figure}

\begin{figure}[t]
    \centering
    \begin{subfigure}[b]{\linewidth}
        \centering
        \includegraphics[width=1\linewidth]{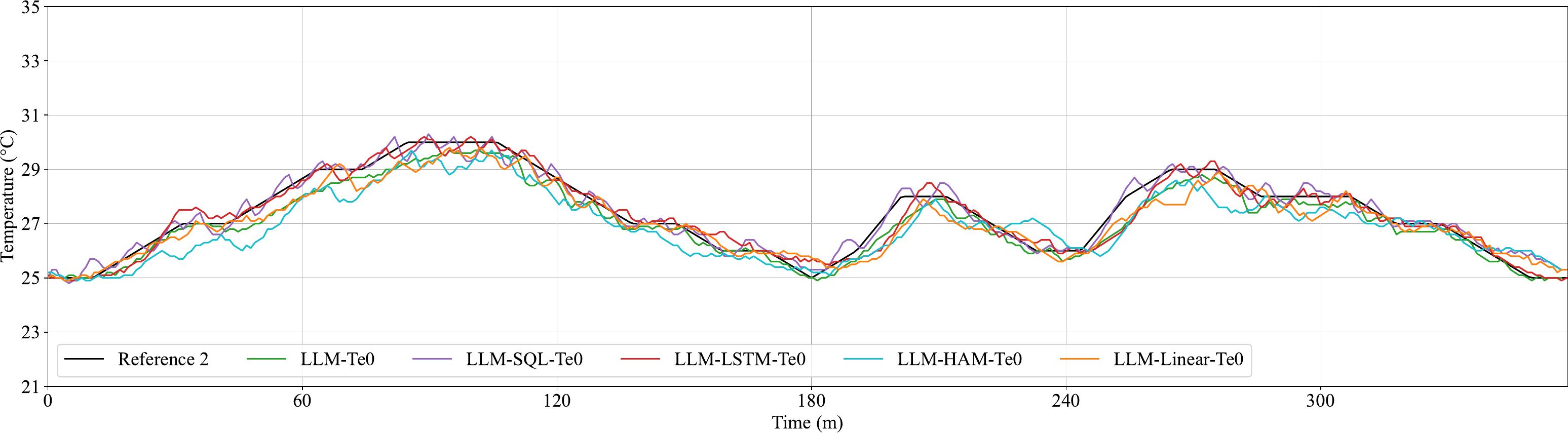}  
        \caption{Temperature}
    \end{subfigure}
    \begin{subfigure}[b]{\linewidth}
        \centering
            \includegraphics[width=1\linewidth]{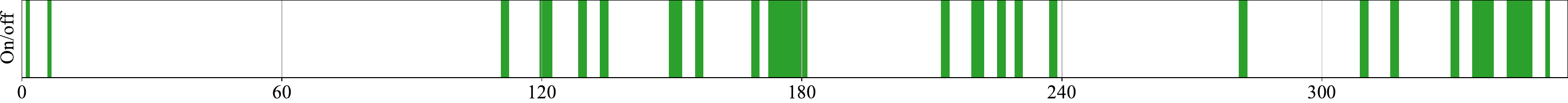}
            \includegraphics[width=1\linewidth]{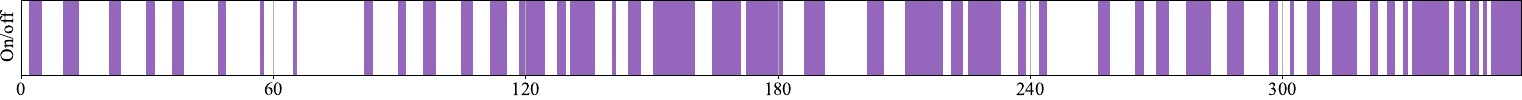}
            \includegraphics[width=1\linewidth]{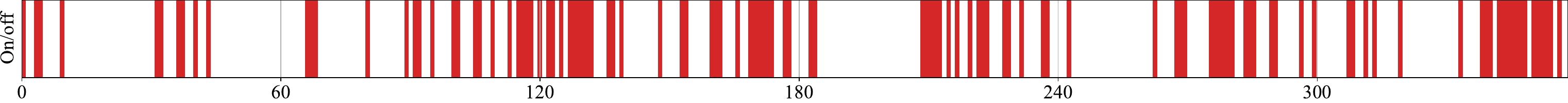}
            \includegraphics[width=1\linewidth]{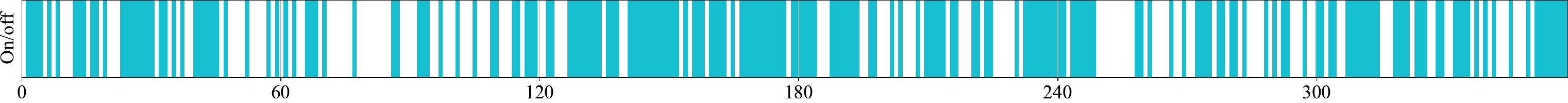}
            \includegraphics[width=1\linewidth]{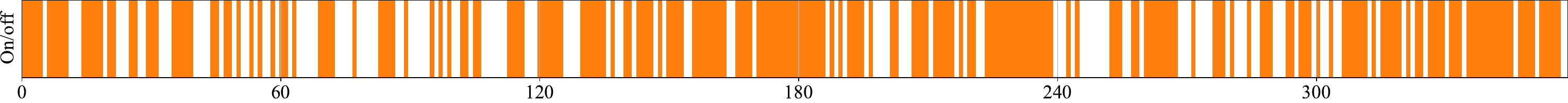}
        \caption{Fan}
    \end{subfigure}
    \begin{subfigure}[b]{\linewidth}
        \centering
            \includegraphics[width=1\linewidth]{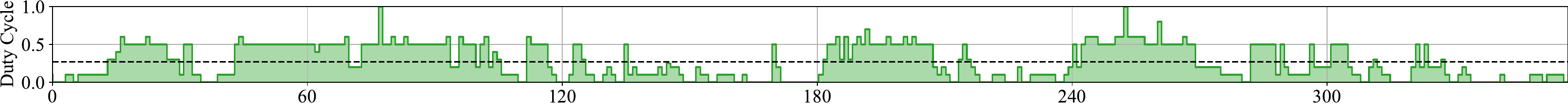}
            \includegraphics[width=1\linewidth]{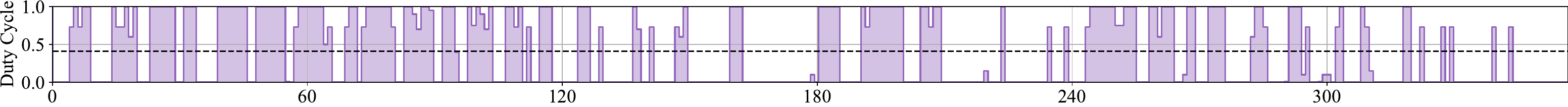}
            \includegraphics[width=1\linewidth]{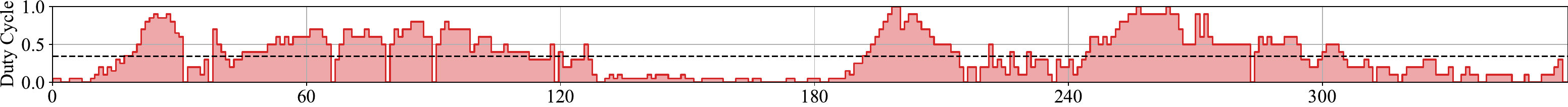}
            \includegraphics[width=1\linewidth]{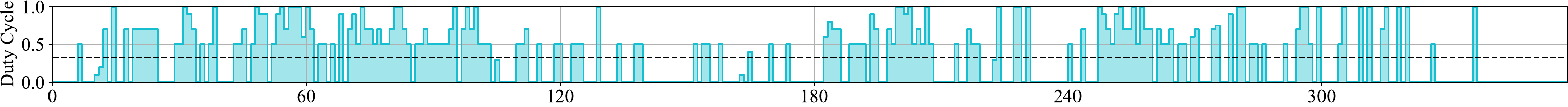}
            \includegraphics[width=1\linewidth]{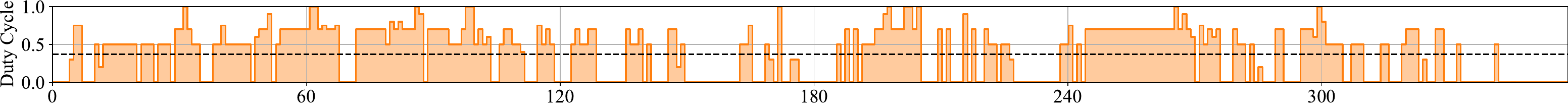}
        \caption{Heater Duty Cycle, average (\dashed)}
    \end{subfigure}
    \caption{Results from running the LLM controllers witout trying to minimize fan usage.}
    \label{fig:results-controller-llm-comparison}
\end{figure}

\begin{figure}[t]
    \centering
    \begin{subfigure}[b]{\linewidth}
        \centering
        \includegraphics[width=1\linewidth]{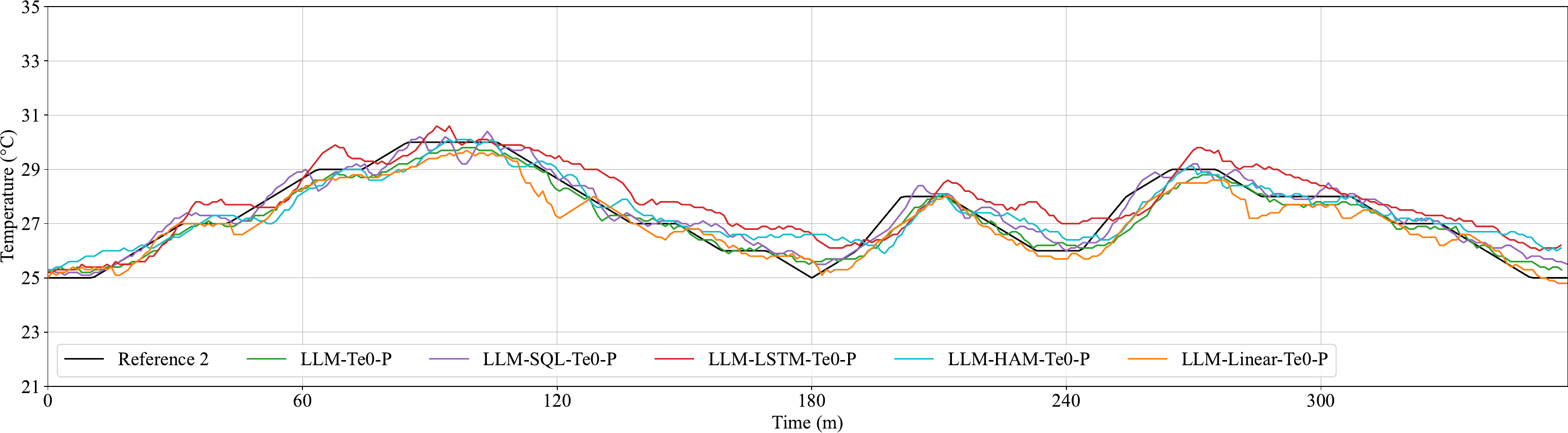}  
        \caption{Temperature}
    \end{subfigure}
    \begin{subfigure}[b]{\linewidth}
            \centering
            \includegraphics[width=1\linewidth]{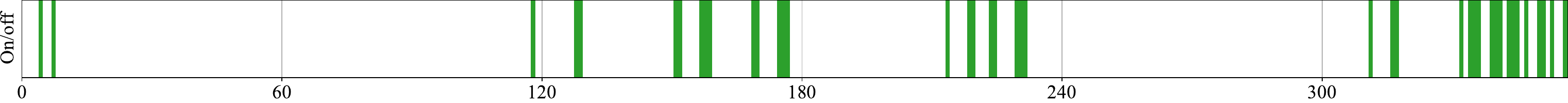}
            \includegraphics[width=1\linewidth]{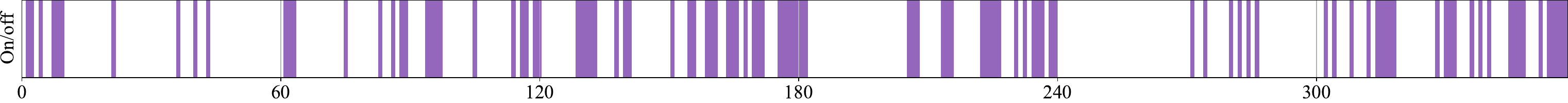}
            \includegraphics[width=1\linewidth]{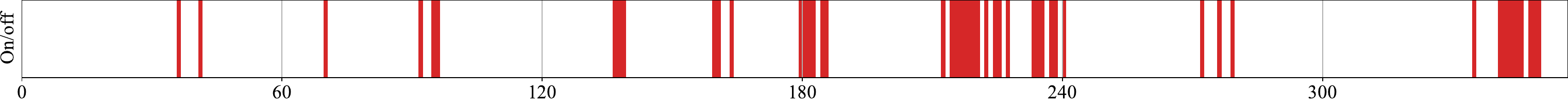}
            \includegraphics[width=1\linewidth]{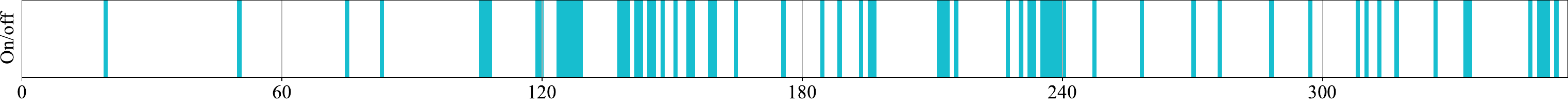}
            \includegraphics[width=1\linewidth]{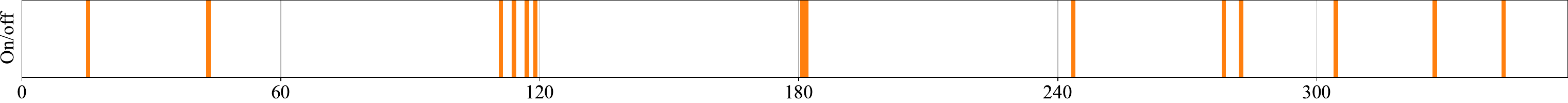}
        \caption{Fan}
    \end{subfigure}
    \begin{subfigure}[b]{\linewidth}
        \centering
            \includegraphics[width=1\linewidth]{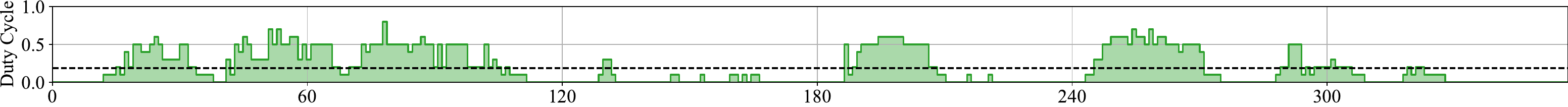}
            \includegraphics[width=1\linewidth]{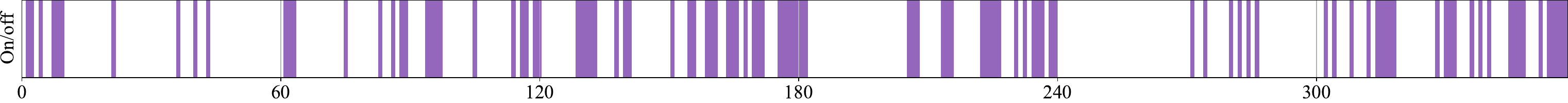}
            \includegraphics[width=1\linewidth]{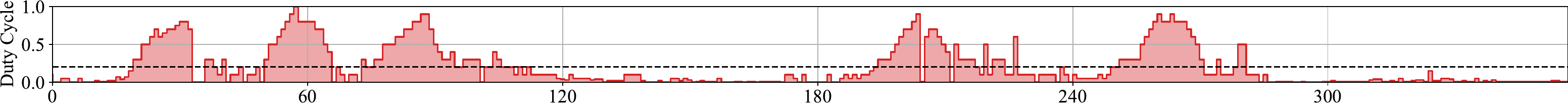}
            \includegraphics[width=1\linewidth]{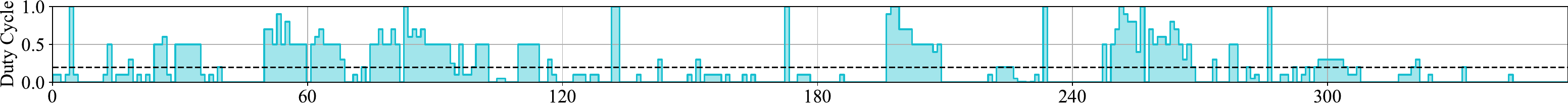}
            \includegraphics[width=1\linewidth]{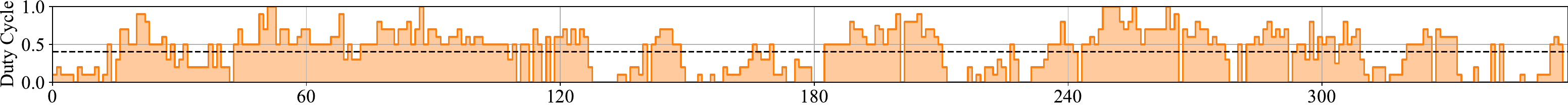}
        \caption{Heater Duty Cycle, average (\dashed)}
    \end{subfigure}
    \caption{Results from running the LLM controllers while trying to minimize the fan usage.}
    \label{fig:results-controller-llm-comparison-penalty}
\end{figure}

\begin{figure}[htbp]
    \centering
    \begin{subfigure}{\linewidth}
        \includegraphics[width=1\linewidth]{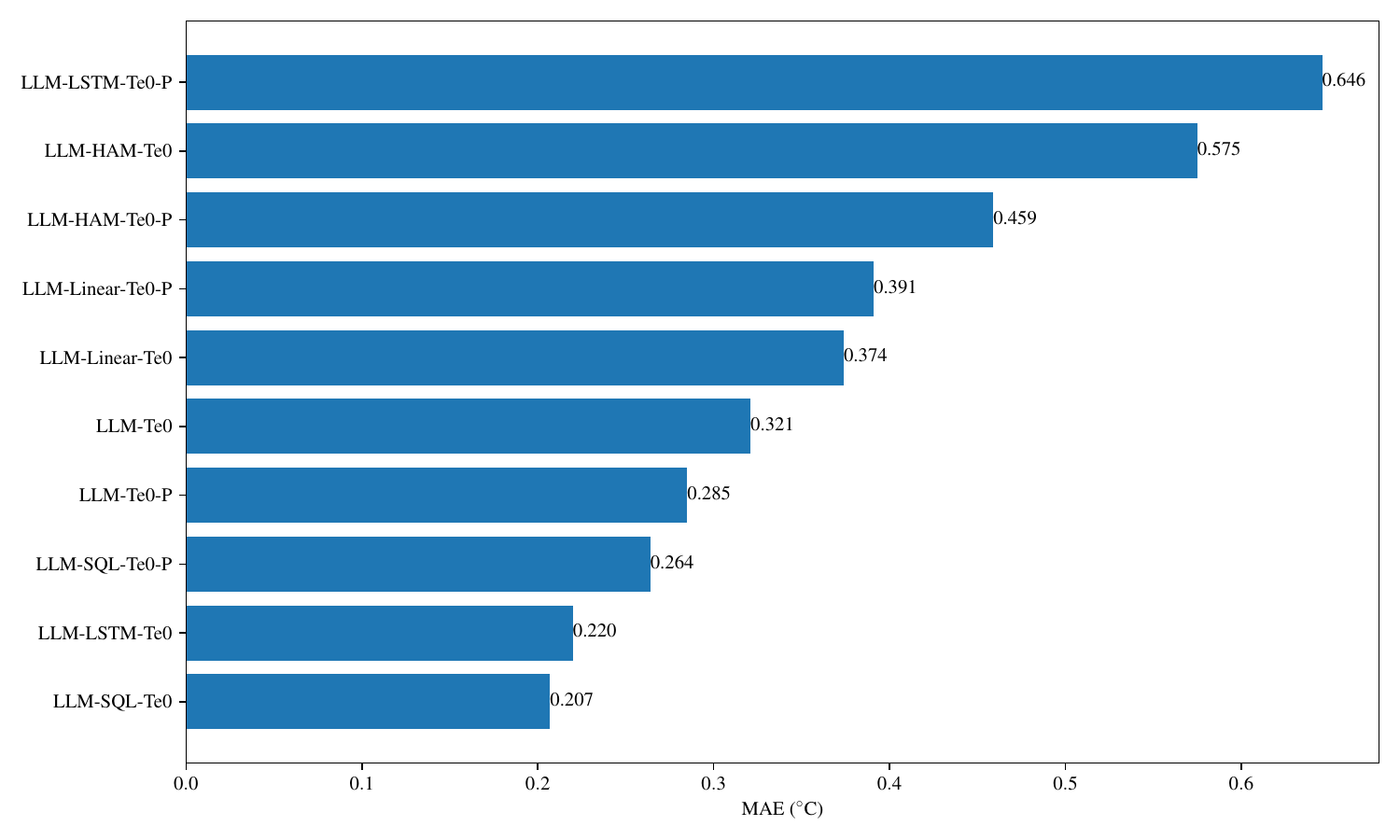}
        \caption{Mean absolute error in tracking temperature}
        \label{subfig:controller_mae_bar}
    \end{subfigure}\\
    \begin{subfigure}{\linewidth}
        \includegraphics[width=1\linewidth]{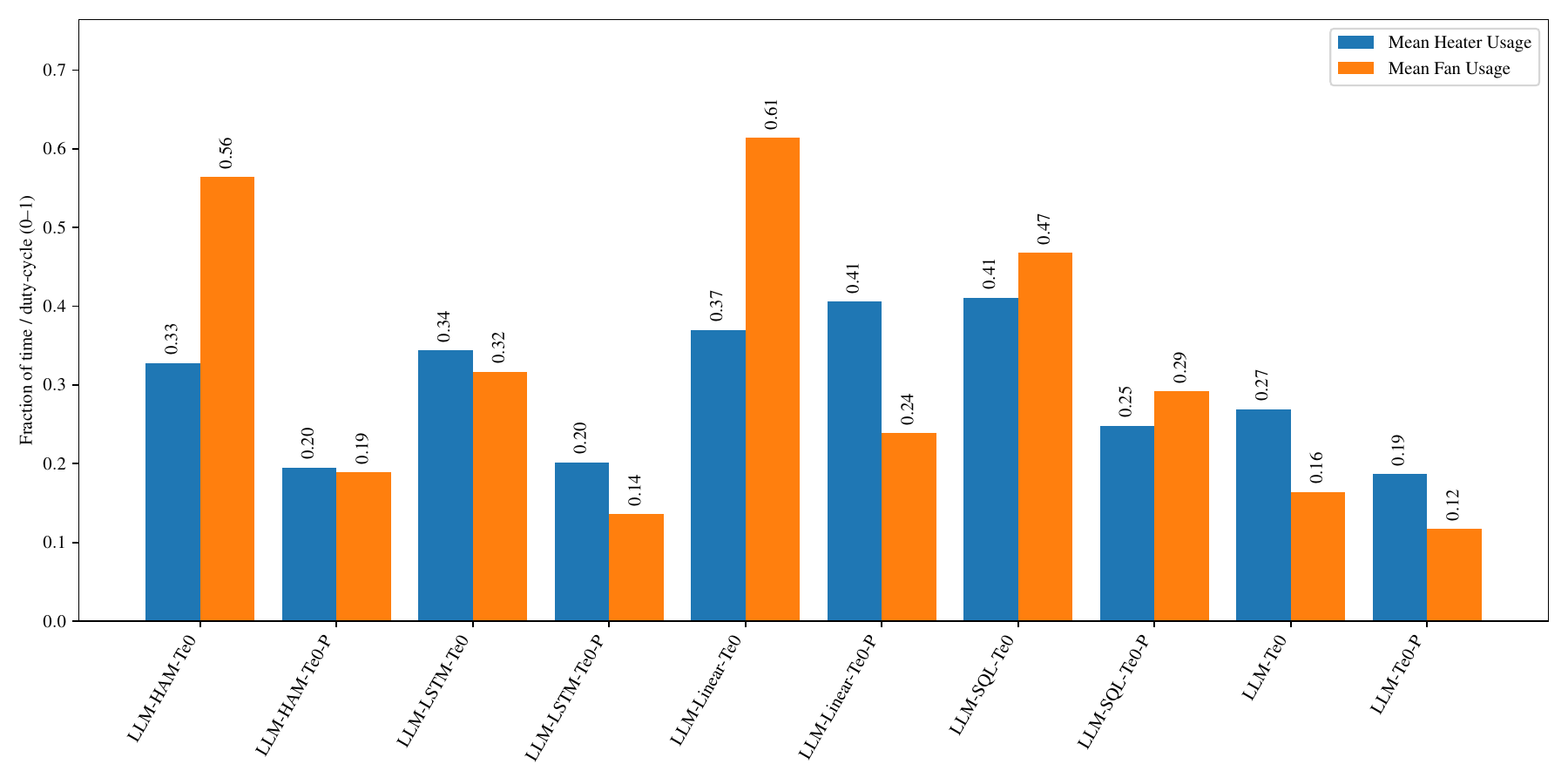}
        \caption{Average actuation usage}
        \label{subfig:controller_usage_grouped}
    \end{subfigure}
    \caption{Performance of different kind of LLM controllers in terms of tracking error and actuaion usage}
    \label{fig:allLLMcomparison}
\end{figure}

This section interprets the controller behaviors shown in Fig.~\ref{fig:results-controller-llm-comparison}, Fig.~\ref{fig:results-controller-llm-comparison-penalty}, and Fig.~\ref{fig:allLLMcomparison}. For the presentation of the results, the controller names follow the convention $\text{Core}[{-}\text{Assistance}] \text{-} \mathrm{Te}\langle \tau \rangle [\text{-P}]$: Core is LLM; Assistance is the tool used (SQL, Linear, LSTM, or HAM; omitted if none). Te$\langle \tau \rangle$ is the LLM creativity (e.g., Te0 deterministic, Te1 creative); and -P indicates the application of the control penalty. In the presentation of the results, we focus on three questions: (i) how well the different LLM variants track the reference temperature, (ii) how they use the actuators (heater and fan), and (iii) how the actuation penalty reshapes their policies. But before we dive into these questions, we present an intercomparison of the different kinds of models used as tools in the current work. Fig.~\ref{fig:modellingerror} presents the comparison. One can clearly see that HAM and LSTM easily outperform both linear and pure PBM models in terms of the mean absolute error. A more detailed comparison of the different models can be found in \cite{rasheed2025hms}. In this work, we confine ourselves to the LSTM and HAM predictive models. 

\subsection{Tracking without actuation penalty}
When no explicit penalty is imposed, all LLM-based controllers are able to follow the reference trajectory, but they do so with clearly different actuation patterns.

\begin{itemize}
    \item \textbf{LLM-Te0} tracks reasonably well and uses the fan sparingly. Its heater decisions are mostly in a small set of values (e.g. 0 or intermediate duty cycles), which leads to a relatively smooth output.
    \item \textbf{LLM-SQL-Te0} benefits from access to historical experiments and tends to reproduce actuator choices that previously worked under similar scenarios. This results in stable behavior, but mostly with more saturated heater actions (close to 0 or 1).
    \item \textbf{Prediction-assisted} variants (LLM-LSTM-Te0, LLM-HAM-Te0, LLM-Linear-Te0) show richer heater dynamics; they attempt more levels because they can evaluate candidates through the predictor. However, this does not automatically yield a lower error. In Fig.~\ref{fig:results-controller-llm-comparison} some of these variants still oscillate once they are close to the target.
\end{itemize}

Overall, Fig.~\ref{fig:results-controller-llm-comparison} shows that a \emph{pure} LLM controller is already viable, and that adding assistance (SQL or a predictor) mostly changes the style of control rather than guaranteeing lower error.

\subsection{Effect of actuation penalty}
When the prompt is modified to activate the penalty, the controllers behave differently.

\begin{itemize}
    \item For \textbf{LLM-Te0} and \textbf{LLM-SQL-Te0}, the penalty has only a modest effect: these variants were already frugal with the fan, so the penalized versions (LLM-Te0-P, LLM-SQL-Te0-P) look similar to the originals.
    \item For \textbf{prediction-assisted} controllers, the effect is stronger. All the variants clearly reduce fan usage, but in periods where the reference temperature decreases, they also show a larger tracking error. This is consistent with the fact that the fan is the main cooling actuator: once it is discouraged, the controller runs out of good actions.
\end{itemize}

Thus, Fig.~\ref{fig:results-controller-llm-comparison-penalty} illustrates that promptable objectives \emph{do} work; fan usage decreases across models—although the cost is sometimes paid in tracking performance, especially for the more ambitious (prediction-based) controllers.

\subsection{Aggregate comparison}
Fig.~\ref{fig:allLLMcomparison} summarizes the behaviour in terms of: (i) mean absolute error (MAE) and (ii) mean heater and fan usage.

\begin{itemize}
    \item On \textbf{MAE} (Fig.~\ref{subfig:controller_mae_bar}), the best values are achieved by the SQL-assisted and LSTM-assisted variants without penalty (e.g. LLM-SQL-Te0, LLM-LSTM-Te0), closely followed by the simple LLM with penalty (LLM-Te0-P). The linear and HAM variants lie in the mid range.
    \item On \textbf{actuator usage} (Fig.~\ref{subfig:controller_usage_grouped}), all ``-P'' controllers reduce fan time as intended; in several cases the heater average is also reduced, indicating that the penalty leads to an overall more economical policy.
    \item There is \textbf{no single controller} that is best on both MAE and actuator economy. Some are more accurate but use more fan; others save actuation but accept a larger error.
\end{itemize}

\subsection{Explaining the actions}
In this section, we present the chain of thinking leading to the recommended actions by the LLM. The actual text related to the thinking and reasoning of LLM can be wordy and vary over different steps; here, we just present some summarized samples. To see the expanded version, we refer the readers to \cite{Ravik2025ill}. We report controller outputs using compact cards: the \emph{Prompt}, the \emph{Tool use} log (queries and calls), a short \emph{Retrieved evidence} (sample), the final \emph{ Controller decision}, and a concise \emph{Brief rationale}. 

\subsubsection{Example A — LLM-SQL}
In this example, the controller was provided with the following prompt and instructed to use an existing database collected during the operation of the setup under model predictive control (MPC). A detailed comparison of MPC, Reinforcement Learning, and LLM can be found in \cite{rasheed2025hms}. The LLM then ran a query, retrieved the relevant data from the database, and used it to recommend an appropriate set of control inputs, along with the rationale behind the recommendation.  
\begin{promptbox}
What should the control values heater\_duty\_cycle and fan\_on be set to in order to maintain a temperature of 27.34C ? The temperature now is 27.1C and the ambient temperature is 22.6C. The first priority is to have minimal usage of the fan when possible, and the second priority is to match the target temperature accurately. Use SQL databse of earlier MPC titled ``MPC control\_penalty2025-03-01T12:41:30" experiment as guide. 
\end{promptbox}

\begin{toolbox}
\textbf{SQL 1}\\[-2mm]
\begin{lstlisting}[style=sqlstyle]
SELECT StartTime, EndTime 
FROM experiments 
WHERE ExperimentID = 'MPC control_penalty2025-03-01T12:41:30';
\end{lstlisting}
\emph{Result:} 2025-03-01 12:45:58 → 2025-03-01 18:44:58

\vspace{2mm}
\textbf{SQL 2}\\[-2mm]
\begin{lstlisting}[style=sqlstyle]
SELECT Temperature, HeaterDutyCycle, FanOn
FROM timeseries_data
WHERE MeasurementTime
BETWEEN `2025-03-01 12:45:58' AND `2025-03-01 18:44:58';
\end{lstlisting}
\end{toolbox}

\begin{evidencebox}
\begin{minipage}{\linewidth}\centering
\begin{tabular}{@{}S[table-format=2.1] S[table-format=1.3] S[table-format=1.0]@{}}
\toprule
{\textbf{Temp (C)}} & {\textbf{HeaterDuty}} & {\textbf{FanOn}}\\
\midrule
25.8 & 0.418 & 0\\
27.3 & 1.000 & 0\\
28.4 & 0.789 & 0\\
29.7 & 1.000 & 0\\
$\cdots$ & $\cdots$ & $\cdots$ \\
30.3 & 0.783 & 0\\
\bottomrule
\end{tabular}
\end{minipage}
\end{evidencebox}

\begin{decisionbox}
\textbf{Heater duty cycle:} \(\approx 0.30\) \quad
\textbf{Fan:} OFF (0)
\end{decisionbox}

\begin{rationalebox}
The heater duty cycle of around 0.3 is chosen to increase the
temperature slightly from 27.1C to 27.34C, as higher duty cycles were used for greater temperature increases in the past experiments. The fan remains off to prevent unnecessary cooling, consistent with historical experiment settings for temperatures within this range
\end{rationalebox}

\subsubsection{Example B — LLM-SQL attempting MPC (failed)}
Similar to Example A, the LLM controller was asked to achieve a certain objective. However, for unexplained reasons, the relevant data could not be retrieved. The controller hallucinated by providing an arbitrary control input (similar to what it had done in the recent past). However, a quick view of the ``Retrieved" evidence and ``Brief rationale" exposes the failure. 

\begin{promptbox}
What should the control values heater\_duty\_cycle and fan\_on be set to in order to maintain a temperature of 31.0C ? The temperature now is 28.4C and the ambient temperature is 22.6C. Use SQL database of earlier MPC titled `MPC control\_penalty2025-03-01T12:41:30' experiment as a guide ?
\end{promptbox}

\begin{toolbox}
\textbf{SQL 1}\\[-2mm]
\begin{lstlisting}[style=sqlstyle]
SELECT StartTime, EndTime 
FROM experiments 
WHERE ExperimentID = 'MPC control_penalty2025-03-01T12:41:30';
\end{lstlisting}
\emph{Result:} 2025-03-01 12:45:58 → 2025-03-01 18:44:58

\vspace{2mm}
\textbf{Intended SQL 2} failed
\end{toolbox}

\begin{evidencebox}
No time-series evidence gathered for the specified window due to early termination.
\end{evidencebox}

\begin{decisionbox}
\textbf{Heater duty cycle:} 0.75 \quad
\textbf{Fan:} ON (1)
\end{decisionbox}

\begin{rationalebox}
Decision cites ``historical data'' but lacks retrieved support because the second query was not run. 
\textbf{Failure mode:} premature stop in tool use \(\rightarrow\) unsupported parameter choice. 
\textbf{Mitigation:} require evidence presence check before emitting actions; re-prompt to fetch relevant rows within the time window.
\end{rationalebox}

\subsubsection{Example C — LLM with LSTM predictor}
Once again, the LLM was tasked with achieving a specific objective with a constraint. However, unlike the previous examples, this time the LSTM predictive model was given as a tool. The LLM used the LSTM model to intelligently evaluate different combinations of control actions and recommend the most optimal combination.

\begin{promptbox}
What should the control values heater\_duty\_cycle and fan\_on be set to in order to maintain a temperature of $30^\circ$C? The temperature now is $27^\circ$C and the ambient temperature is $23^\circ$C. The first priority is to have minimal usage of the fan when possible, and the second priority is to match the target temperature accurately. It is important that the temperature in the greenhouse matches the target temperature quite well. Use LSTM.  
\end{promptbox}

\begin{toolbox}
\textbf{Predictor calls:} Evaluate constant sequences using LSTM\\
(Heater=0, Fan=0), (Heater=1, Fan=0), (Heater=0.5, Fan=0), (Heater=0.8, Fan=0)\\
Predicted end-of-horizon temperature using LSTM: 26.80, 30.08, 28.43, 28.93 respectively.
\end{toolbox}

\begin{evidencebox}
(Heater=0.8,Fan=0) gave the temperature closest to 30. 
\end{evidencebox}

\begin{decisionbox}
\textbf{Heater duty cycle:} 0.80 \quad \textbf{Fan:} OFF
\end{decisionbox}

\begin{rationalebox}
(Heater=1, Fan=0) can create overshoot requiring cooling in the next step so chose (Heater=0.8, Fan=0).
\end{rationalebox}

\subsubsection{Example D — LLM with HAM predictor}
This is simiar to the previous example except that the LLM was asked to use the HAM model. 
\begin{promptbox}
What should the control values heater\_duty\_cycle and fan\_on be set to in order to maintain a temperature of $30^\circ$C? The temperature now is $27^\circ$C and the ambient temperature is $23^\circ$C. The first priority is to have minimal usage of the fan when possible, and the second priority is to match the target temperature accurately. Use HAM.
\end{promptbox}

\begin{toolbox}
\textbf{Predictor calls:} Evaluate constant sequences using HAM\\
(Heater=0, Fan=0), (Heater=1, Fan=0), (Heater=0.5, Fan=0), (Heater=0.8, Fan=0)\\
Predicted end-of-horizon temperature using HAM: 27.60, 28.97, 28.43, 28.83 respectively.
\end{toolbox}

\begin{evidencebox}
All no-fan sequences raise \(T\) modestly; Heater=1.0 gives the largest increase but still below $30^\circ$C over 10 steps.
\end{evidencebox}

\begin{decisionbox}
\textbf{Heater duty cycle:} 1.00 \quad
\textbf{Fan:} OFF
\end{decisionbox}

\begin{rationalebox}
The predictions show that with the heater ON and the fan OFF,
the temperature will gradually increase towards the target of $30^{\circ}$C. Keeping the fan off aligns with the priority of minimal fan usage, and the heater will help in reaching the target temperature more effectively.
\end{rationalebox}

\subsection{Discussion}
\label{subsec:discussion}

The experiments show that LLM-based controllers form a spectrum rather than a single best solution. Prompt-only controllers are the most stable and least sensitive to changes in the prompt or operating conditions; however, they do not always achieve the lowest tracking error. SQL-assisted controllers are a strong middle ground because access to historical episodes allows the LLM to imitate previously successful behavior and keeps the MAE competitively low, even when an actuation penalty is introduced. Prediction-assisted controllers (Linear, LSTM, HAM) provide the LLM with the richest basis for choice and, therefore, adapt most strongly when the objective changes (for example, when fan usage is to be minimized). At the same time, these variants are also the ones that degrade the most when the prompt removes or discourages a particular actuation, illustrating that more assistance also means a greater sensitivity to constraints.

A second observation is that there is no single configuration that is simultaneously optimal on tracking accuracy and actuator economy. Some controllers follow the reference closely but use more fan or heater time; others achieve lower actuation but accept a larger MAE. This suggests that LLM-for-control should be treated as a \emph{family} of controllers that can be selected or switched at runtime depending on what matters most at that moment (accuracy, energy, comfort, or actuator wear).

Finally, the \emph{Prompt→Evidence→Decision} presentation used in this paper is not only convenient for reporting but also exposes concrete failure modes. By listing the prompt, the tool calls, the retrieved evidence, and the final decision together with a brief rationale, we can immediately see when the LLM stopped the tool chain too early, when it acted without sufficient evidence, or when it ignored an instruction in the prompt. This makes LLM-in-the-loop control more auditable and gives a practical way to add guardrails (for example, “do not act unless at least one row was retrieved” or “re-query if the time window is empty”) before deployment.

\section{Conclusion}
\label{sec:conclusion}
This paper demonstrates how LLMs can be applied to control tasks when provided with task context, explicit objectives, and, optionally, tool-based assistance. The work showed that a plain, prompt-driven LLM can already produce sensible control actions, lowering the entry barrier compared to hand-crafted controllers that require modeling and tuning expertise. Adding assistance does not automatically improve accuracy, but it does make the controller more purposeful. With SQL, the LLM can reuse successful past behavior, and with prediction models, it can evaluate multiple candidate action sequences before selecting one. A key observation is that the control style can be changed through prompting. Simple changes to the instructions, such as asking to minimize actuator usage, lead to consistent reductions across several variants. This is attractive to operators who might prefer to express operational preferences in natural language rather than redesigning cost functions.

At the same time, the study showed that no single LLM configuration is best in all metrics. Tool augmented variants can become sensitive to constraints. Penalty based runs can save energy but accept a higher tracking error. Prompt-only runs remain the most robust, but they are not always the most accurate. This suggests that LLM-for-control should be treated as a family of controllers rather than a single solution, with the option to switch or select a variant according to current objectives, actuator limits, and data quality.

Finally, because LLM decisions are produced in natural language and can be logged together with queries and retrieved data, the approach supports transparent, auditable control actions. 

Future work should focus on automating the choice of assistance (when to call SQL and when to call a predictor), using local lightweight models that can be tuned to a specific application, and enabling the LLM to employ the prediction model for longer-horizon planning in a way that resembles MPC. It should also integrate hard safety checks around the LLM output

\section*{DECLARATION OF GENERATIVE AI AND AI-ASSISTED TECHNOLOGIES IN THE WRITING PROCESS} During the preparation of this work, the author(s) used ChatGPT-5 in order to improve the readability of this article.. After using this tool/service, the author(s) reviewed and edited the content as needed and take full responsibility for the content of the publication.
\bibliography{ifacconf}  
\end{document}